\documentclass[a4paper,11pt]{article}
\usepackage{amsfonts}
\usepackage{amssymb}
\begin{document}
\title{Maxwell-Boltzmann type Hawking radiation}
\author{Youngsub Yoon \\\emph{Dunsan-ro 201, Seo-gu} \\\emph{Daejeon 35245, Korea}}
\maketitle

\begin{abstract}
Twenty years ago, Rovelli proposed that the degeneracy of black hole (i.e. the exponential of the Bekenstein-Hawking entropy) is given by the number of ways the black hole horizon area can be expressed as a sum of unit areas. However, when counting the sum, one should treat the area quanta on the black hole horizon as distinguishable. This distinguishability of area quanta is noted in Rovelli's paper. Building on this idea, we derive that the Hawking radiation spectrum is not given by Planck radiation spectrum (i.e., Bose-Einstein distribution) but given by Maxwell-Boltzmann distribution.
\end{abstract}

\section{Background}
According to loop quantum gravity \cite{Discreteness, General, Ashtekar}, the eigenvalues of the area operator are quantized, and the black hole area, as much as any area, is the sum of these eigenvalues. For example, let us say that we have the following area eigenvalues (i.e., the unit areas):

\begin{equation}
A_i=A_1, A_2, A_3, A_4, A_5, A_6....
\end{equation}
Then, the black hole area $A$ must be given by the following formula:

\begin{equation}
A=\sum_i N_i A_i\label{NjAj},
\end{equation}
where the $N_i$s are non-negative integers. Here, we can regard the black hole as having $\sum N_i$ partitions, each of which has one of the $A_i$ as its area.

Using the discreteness of the area, in 1996, Rovelli \cite{Rovelli} proposed that the degeneracy of black hole (i.e., the exponential of the Bekenstein-Hawking entropy) is given by the number of ways the black hole horizon area can be expressed as a sum of unit areas. However, when counting the sum, one should treat the area quanta on the black hole horizon as distinguishable. This distinguishability of area quanta is noted in Rovelli's paper: one should treat quanta as distinguishable if they have fixed locations. Indeed, area quanta have fixed locations on the black hole horizon.

In \cite{QCHawking} we derived a selection rule for the Hawking radiation using elementary derivation of Bose-Einstein distribution. The selection rule was that upon an emission of a photon by a black hole, the horizon area $A$ decreases by a unit area. In other words,

\begin{equation}
\Delta A=-A_i
\end{equation}
As the Bekenstein-Hawking entropy is given by $S=kA/4$, and we know $\Delta Q=T\Delta S$, the energy decrease is given by
\begin{equation}
\Delta Q=-kT\frac{A_i}{4}
\end{equation}
Since this energy must be equal to the energy of photon emitted (i.e., $\Delta Q=-hf$) the frequency of the photon emitted during the Hawking radiation is given by
\begin{equation}
f_i=\frac{kT}{h}\frac{A_i}{4}\label{f=kT/hA/4}
\end{equation}

In the next section, we explain why we should use the Maxwell-Boltzmann distribution instead of the Bose-Einstein distribution for the Hawking radiation.

\section{Maxwell-Boltzmann distribution}
This section closely uses the method presented in the famous quantum mechanics textbook by Griffiths \cite{Griffiths}:

Let us say that the unit areas $A_1, A_2, A_3,\cdots$ have degeneracies $d_1,d_2,d_3,\cdots$. Suppose we have a black hole with area $A$ which satisfies $A=\sum_i N_i A_i$ as explained before. For a given configuration ($N_i=N_1,N_2,N_3,\cdots$), how many different ways can this be achieved?

Now, recall that the area quanta is distinguishable. Then, the answer is given by
\begin{equation}
Q=N!\prod_{i=1}^{\infty}\frac{d_i^{N_i}}{N_i!}
\end{equation}
where $N=\sum_i N_i$. Recall that we also have the following condition
\begin{equation}
A=\sum_{n=1}^{\infty}N_iA_i
\end{equation}
To find the most probable configuration $(N_1,N_2,N_3,\cdots)$, we need to maximize $\ln Q$ as follows:
\begin{equation}
G\equiv \ln
Q+\alpha\left[A-\sum_{n=1}^{\infty}N_iA_i\right],\label{G}
\end{equation}
where $G$ is to be maximized and $\alpha$ is a Lagrange
multiplier. Let us maximize it by differentiating with respect to $N_i$. First, note
\begin{eqnarray}
\ln Q(\cdots,N_{i-1}, N_i,N_{i+1},\cdots)=N_i\ln d_i+\ln N!~~~~~~~~~~~~~~~~~~~~~~~~~~~~~~~~~~~~\nonumber\\-(\cdots+\ln N_{i-1}!+\ln N_i!+\ln N_{i+1}!+\cdots)
\end{eqnarray}
Then, using $N-1=N_1+\cdots+N_{i-1}+N_i-1+N_{i+1}+\cdots$, we have
\begin{eqnarray}
\ln Q(\cdots,N_{i-1}, N_i-1,N_{i+1},\cdots)=(N_i-1)\ln d_i+\ln (N-1)!-(\cdots+\ln N_{i-1}!\nonumber\\+\ln (N_i-1)!+\ln N_{i+1}!+\cdots)~~~~~~~~~~
\end{eqnarray}
Using these, we have
\begin{eqnarray}
0=\frac{\partial G}{\partial N_i}=\ln Q(\cdots,N_{i-1}, N_i,N_{i+1},\cdots)-\ln Q(\cdots,N_{i-1}, N_i-1,N_{i+1},\cdots)-\alpha A_i\nonumber\\=
\ln d_i+(\ln N!-\ln (N-1)!)-(\ln N_i!-\ln (N_i-1)!)-\alpha A_i~~~~~~
\end{eqnarray}

The conclusion is
\begin{equation}
\frac{N_i}{N}=\frac{d_i}{e^{\alpha A_i}}\label{Ni}
\end{equation}
On the other hand, if Hawking radiation were given by Bose-Einstein distribution, we know that the Hawking radiation for large photon frequency is given by
\begin{equation}
N_i=\frac{d_i}{e^{hf_i/(kT)}-1}
\end{equation}
This expression must reduce to (\ref{Ni}) for large $f_i$. Using (\ref{f=kT/hA/4}) we conclude $\alpha=1/4$. Therefore, (\ref{Ni}) becomes
\begin{equation}
\frac{N_i}{N}=\frac{d_i}{e^{hf_i/(kT)}}=\frac{d_i}{e^{A_i/4}}
\end{equation}
We can check that our calculation is indeed correct. Summing the both sides, we get
\begin{equation}
\sum_i \frac{N_i}{N}=\sum_i d_i e^{-A_i/4}
\end{equation} 
The left-hand side is 1 by the definition of $N$. The right-hand side is also 1 by Domagala-Lewandowski-Meissner formula \cite{Domagala, Meissner}.

\section{Discussions and Conclusions}
Even though our result that the Hawking radiation follows the Maxwell-Boltzmann distribution is different from the currently accepted Bose-Einstein distribution, it may not be easy to experimentally confirm this even if Hawking radiation is observed, as $e^{A_i/4}$ is much bigger than 1. For example, for $A_1$, the smallest unit area, $e^{A_1/4}$ is about 85 \cite{predictions}. 
\pagebreak


\begin{thebibliography}{9}

\bibitem{Discreteness}
C.~Rovelli, L.~Smolin,
``Discreteness of area and volume in quantum gravity,''
Nucl.\ Phys.\  {\bf B442}, 593-622 (1995).
[gr-qc/9411005].

\bibitem{General}
S.~Frittelli, L.~Lehner, C.~Rovelli,
``The Complete spectrum of the area from recoupling theory in loop quantum gravity,''
Class.\ Quant.\ Grav.\  {\bf 13}, 2921-2932 (1996).
[gr-qc/9608043].

\bibitem{Ashtekar}
A.~Ashtekar, J.~Lewandowski,
``Quantum theory of geometry. 1: Area operators,''
Class.\ Quant.\ Grav.\  {\bf 14}, A55-A82 (1997).
[gr-qc/9602046].


\bibitem{Rovelli}
C.~Rovelli,
``Black hole entropy from loop quantum gravity,''
Phys.\ Rev.\ Lett.\  {\bf 77}, 3288-3291 (1996).
[gr-qc/9603063].

\bibitem{QCHawking} 
Y.~Yoon,
``Quantum corrections to the Hawking radiation spectrum,''
J.\ Korean Phys.\ Soc.\  {\bf 68}, no. 6, 730 (2016)
doi:10.3938/jkps.68.730
[arXiv:1210.8355 [gr-qc]].
%%CITATION = doi:10.3938/jkps.68.730;%%
%6 citations counted in INSPIRE as of 01 Jun 2016

\bibitem{Griffiths} 
D.~J.~Griffiths,
``Introduction to quantum mechanics,'' Prentice Hall, 2005.


\bibitem{Domagala}
M.~Domagala, J.~Lewandowski,
``Black hole entropy from quantum geometry,''
Class.\ Quant.\ Grav.\  {\bf 21}, 5233-5244 (2004).
[gr-qc/0407051].

\bibitem{Meissner}
K.~A.~Meissner,
``Black hole entropy in loop quantum gravity,''
Class.\ Quant.\ Grav.\  {\bf 21}, 5245-5252 (2004).
[gr-qc/0407052].

\bibitem{predictions} 
 B.~Kong and Y.~Yoon,
  ``Black hole entropy and Hawking radiation spectrum predictions without Immirzi parameter and Hawking radiation of single-partition black hole,''
 J.\ Korean Phys.\ Soc.\  {\bf 68}, no. 6, 735 (2016)
 doi:10.3938/jkps.68.735
 [arXiv:0910.2755 [gr-qc]].
 %%CITATION = doi:10.3938/jkps.68.735;%%
 %5 citations counted in INSPIRE as of 01 Jun 2016
 

\end{thebibliography}
\end{document}